\documentclass{emulateapj}

\def\kmsec{\mbox{km~s$^{\rm -1}$}}
\def\logg{\mbox{log~{\it g}}}
\def\teff{$T$}
\def\vt{\mbox{$v_{\rm t}$}}
\def\rpro{\mbox{$r$-process}}
\def\spro{\mbox{$s$-process}}

\def\loggf{$\log$($gf$)}

\shorttitle{Stars Lacking Neutron-Capture Elements}
\shortauthors{I.~U.\ Roederer}
\slugcomment{Accepted for publication in the Astronomical Journal}

\begin{document}

\title{Are There Any Stars Lacking Neutron-Capture Elements? \\
Evidence from Strontium and Barium}

\author{Ian U.\ Roederer}
\affil{Carnegie Observatories,
813 Santa Barbara Street, Pasadena, CA 91101 USA; 
iur@obs.carnegiescience.edu}

\begin{abstract}

The cosmic dispersion in the abundances of
the heavy elements strontium and barium in halo stars
is well known.
Strontium and barium are detected in most cool, metal-poor giants,
but are these elements always detectable?
To identify stars that could be considered probable candidates 
for lacking these elements, I examine the 
stellar abundance data available in the literature for 
1148 field stars and 
226 stars in dwarf galaxies, 
776 of which have metallicities lower than [Fe/H]~$< -$2.0.
Strontium or barium have been detected in all
field, globular cluster, and dwarf galaxy environments studied.
All upper limits are consistent with
the lowest detected ratios
of [Sr/H] and [Ba/H].
The frequent appearance of these elements
raises the intriguing prospect that
at least one kind of neutron-capture reaction
operates as often as
the nucleosynthesis mechanisms that produce lighter elements,
like magnesium, calcium, or iron,
although the yields of heavy elements may be more variable.

\end{abstract}

\keywords{
galaxies: dwarf ---
globular clusters: general ---
nuclear reactions, nucleosynthesis, abundances ---
stars: abundances ---
stars: Population II
}

\section{Introduction}
\label{intro}

It is now well established that there exists a
cosmic star-to-star dispersion in the abundances of
the heavy elements barium and strontium 
in halo field stars.
Early hints of this effect appeared in the work of
\citet{gilroy88} and \citet{magain89}, 
and subsequent work by \citet{ryan91,ryan96}, 
\citet{mcwilliam95}, \citet{mcwilliam98}, \citet{burris00},
and others confirmed and extended
it to stars with metallicities 10,000 times lower than the Sun.
These studies have repeatedly demonstrated that 
the dispersion cannot be a consequence of uncertainties in
the abundance analysis techniques or other observational error.
The lighter elements show more constant ratios with dispersions
mostly consistent with observational error only 
(e.g., \citealt{arnone05}).
The contrast between the light (e.g., magnesium) and heavy elements
(e.g., strontium, barium, or europium) is
apparent in the small dispersion of
[Mg/Fe] ratios and the large dispersion of [Eu/Fe] ratios at low metallicity, 
as shown by Figure~14 in \citet{sneden08}.

Many surveys,
including those of
\citet{bond70,bond80},
\citet{bidelman73},
Beers, Preston, \& Shectman (\citeyear{beers85}, \citeyear{beers92}),
\citet{frebel06},
\citet{christlieb08}, 
and the Sloan Extension for Galactic Understanding 
and Exploration \citep{yanny09}
have been conducted
to identify FGK-type stars that contain
weak absorption lines from metals such as calcium or iron.
Presumably such stars are among the oldest low-mass stars in the Universe,
but to date no star has been found to be lacking 
magnesium, calcium, or iron.
The elements heavier than the iron group ($Z >$~30 or so) are 
thought to be produced in different ways than the lighter elements are,
likely by neutron-capture or charged-particle reactions.
But, like the lighter elements, are the heavy ones always present?

The purpose of this paper is to examine the 
stellar abundance data available in the literature to 
consider whether any stars are currently known 
that may be considered probable candidates for 
lacking elements heavier than the iron group altogether.
The specific nucleosynthetic origins of 
these elements are no doubt interesting, 
but instead 
I will focus on the more general question of detectability.
\citet{roederer10b} considered how
often Sr~\textsc{ii} or Ba~\textsc{ii} lines had been detected in one
large snapshot survey of metal-poor stars, that of \citet{barklem05},
who studied 253~stars.
For the 34~coolest stars in their sample,
those with $T <$~4800~K, 
Sr~\textsc{ii} and Ba~\textsc{ii} lines were detected
in every one.
In this paper, I reexamine this
question by considering an expanded sample of
field, globular cluster, and dwarf galaxy stars.

I adopt the standard definition of elemental ratios
throughout this work.
For elements X and Y, the logarithmic abundance ratio relative to the
solar ratio is defined as
[X/Y]~$\equiv \log_{10} (N_{\rm X}/N_{\rm Y}) -
\log_{10} (N_{\rm X}/N_{\rm Y})_{\odot}$.
These ratios always
indicate the total elemental abundance 
after ionization corrections have been applied.
For stellar metallicity, I adopt the iron abundance as derived
from Fe~\textsc{ii} lines, when available, and otherwise I resort
to the [Fe/H] values reported in the literature.

\section{Strontium and Barium}
\label{srba}

Strontium (Sr, $Z =$~38) and barium (Ba, $Z =$~56)
are the two elements heavier than the iron group
best-suited for this investigation.
Sr~\textsc{ii} and Ba~\textsc{ii} are the dominant
species of these elements in late-type stellar atmospheres
because of their low first ionization potentials,
5.69 and 5.21~eV, respectively.
Both elements are members of the alkaline earth metals along with
magnesium (Mg, $Z =$~12) and calcium (Ca, $Z =$~20).
These ions all have a single valence s electron in the ground state.
The first excited p state has $J =$~3/2 or 1/2, 
giving rise to a strong resonance doublet feature.
These transitions correspond to the well-known
Mg~\textsc{ii} doublet at 2795 and 2802\,\AA\ 
(connecting the 2p$^{6}$3s $^{2}$S$_{1/2}$ and 2p$^{6}$3p $^{2}$P$^{\rm o}$ 
terms)
and the
Ca~\textsc{ii} doublet at 3933 and 3968\,\AA\
(3p$^{6}$4s $^{2}$S$_{1/2}$ to 3p$^{6}$4p $^{2}$P$^{\rm o}$).
These absorption lines frequently rank among the strongest
lines observed in late-type stars and 
galaxies whose integrated light is dominated by
old stellar populations.
The 
Sr~\textsc{ii} doublet at 4077 and 4215\,\AA\ 
(4p$^{6}$5s $^{2}$S$_{1/2}$ to 4p$^{6}$5p $^{2}$P$^{\rm o}$) and the 
Ba~\textsc{ii} doublet at 4554 and 4934\,\AA\
(5p$^{6}$6s $^{2}$S$_{1/2}$ to 5p$^{6}$6p $^{2}$P$^{\rm o}$)
are their analogs.
Since strontium and barium frequently rank among the more
abundant elements heavier than the iron group,
these spectral lines are
the most readily available heavy element transitions
in the optical, near ultraviolet, and near infrared regions 
of stellar spectra.

The proportionally large abundance of barium 
makes it a better choice for comparison
than europium ($Z =$~63).
Europium is commonly used as a tracer of nucleosynthesis
by the rapid neutron-capture process (\rpro).
A large fraction 
of its solar system abundance ($>$~91\%; e.g., 
\citealt{cameron82,burris00,bisterzo11})
is attributed to \rpro\ nucleosynthesis.
In the solar system, barium is $\approx$~46~times 
more abundant than europium.
In metal-poor halo stars strongly enriched by \rpro\ material,
like \mbox{CS~22892--052}, barium is 
approximately nine times more abundant than europium
\citep{sneden03a}.
Metal-poor stars with a deficiency 
of elements heavier than barium, like \mbox{HD~122563},
have similar ratios of barium to europium,
$\approx$~10--15 (e.g., \citealt{honda06,roederer12}).
Metal-poor stars highly enriched by material
produced by slow neutron-capture reactions (the \spro)
frequently show barium to europium ratios
of several hundred or more
(e.g., \citealt{aoki02b}).
Barium is always considerably more
abundant than europium, so it should be 
detectable more often than europium
if the abundances of both elements are low.

In most metal-poor stars, the [Ba/Eu] ratio is low
and close to the ratios found in
\mbox{CS~22892--052} and \mbox{HD~122563}
(e.g., \citealt{gilroy88,mcwilliam98}).
This indicates that even the barium in these stars
owes its origin mainly to some kind of \rpro\ nucleosynthesis.
The strontium may owe its origins to several
mechanisms in addition to \rpro\ nucleosynthesis
(e.g., \citealt{travaglio04}).
I emphasize that the results of the present study
do not require that either of these 
interpretations holds true.

\section{Literature Data}
\label{data}

\begin{figure*}
\begin{center}
\includegraphics[angle=0,width=5.5in]{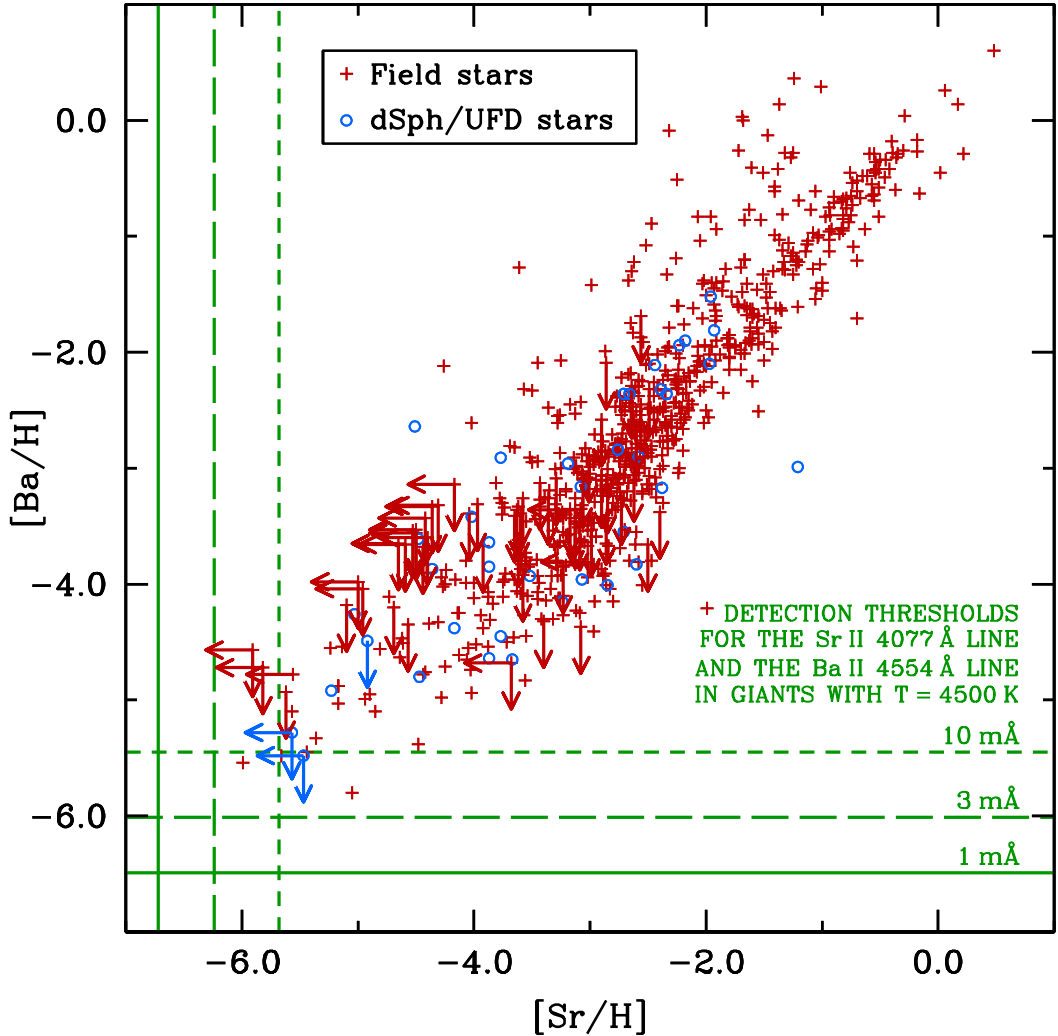}
\end{center}
\caption{
\label{srbaplot}
\scriptsize
[Sr/H] and [Ba/H] ratios in field stars and dwarf galaxies.
Approximate equivalent width detection thresholds 
for the Sr~\textsc{ii} 4077\,\AA\ line 
and the Ba~\textsc{ii} 4554\,\AA\ line
in the atmosphere of a cool giant, assuming LTE,
are shown by the sets of green curves.
The data for the field stars 
are taken from the following sources:
\citet{andrievsky11},
\citet{aoki02,aoki04,aoki05,aoki06,aoki09,aoki10,aoki12},
\citet{bai04},
\citet{barklem05},
\citet{bonifacio09},
\citet{burris00},
\citet{caffau12},
\citet{carretta02},
\citet{christlieb04},
\citet{cohen04,cohen06,cohen08},
\citet{depagne00},
\citet{francois07},
\citet{fulbright00},
\citet{gratton94},
\citet{hansen11},
\citet{hollek11},
\citet{honda04,honda11},
\citet{ishigaki10},
\citet{ito09},
\citet{ivans03},
\citet{johnson02},
\citet{johnson02b},
\citet{jonsell05},
\citet{lai07,lai08,lai09},
\citet{mcwilliam95},
\citet{mcwilliam98},
\citet{mishenina01},
\citet{nissen11},
\citet{norris00,norris01,norris07},
\citet{preston00,preston01},
\citet{preston06},
\citet{roederer10a},
\citet{ryan91,ryan96,ryan98},
\citet{sivarani06},
\citet{sneden03},
\citet{spite00},
\citet{stephens02}, and
\citet{yong12}.
The data for the dSph and UFD
stars are taken from the following sources:
\citet{aoki07,aoki09b},
\citet{cohen09,cohen10},
\citet{cohen12},
\citet{feltzing09},
\citet{frebel10a,frebel10b},
\citet{fulbright04},
\citet{geisler05},
\citet{honda11b},
\citet{kirby12},
\citet{koch08},
\citet{lemasle12},
\citet{letarte10},
\citet{norris10a,norris10b},
\citet{sbordone07},
\citet{shetrone01,shetrone03},
\citet{simon10},
\citet{tafelmeyer10}, and
\citet{venn12}.
 }
\end{figure*}
\normalsize

I have compiled a sample of 1148~field stars and 
226~stars in dwarf spheroidal (dSph) or ultra-faint dwarf (UFD) galaxies
from the literature with reported detections or upper limits 
on the strontium or barium abundance.
Of these, 728 (39) field (dSph and UFD) stars have reported 
detections or upper limits for both strontium and barium,
318 (187) stars have reported
detections or upper limits for barium only, and
102 (0) stars have reported
detections or upper limits for strontium only.
A total of 707 (69) of these stars have [Fe/H]~$< -$2.0.
The majority of the field stars in this compilation were
originally identified as having weak metal lines
in the spectroscopic surveys listed in Section~\ref{intro} or
as high proper motion stars in the surveys of 
Giclas, Burnham, \& Thomas (\citeyear{giclas71}, \citeyear{giclas78}).
All of the abundances considered have been
derived from spectra with moderately high resolution
($R \equiv \lambda/\Delta\lambda \sim$~15,000 or better).
The complete list of 54~studies of field stars and 23~studies
of dSph and UFD galaxy stars is given in the caption to Figure~\ref{srbaplot}.
Additional comments on a few stars can be found in Appendix~\ref{appendix}.

This compilation is surely not complete for the highest metallicities
or for stars with high levels of heavy elements.
For example, stars strongly enriched by the \rpro\ or \spro\
are underrepresented.
These are not the stars of interest here.
Sr~\textsc{ii} and Ba~\textsc{ii} lines
are always easily detectable in these stars,
and the lines are often saturated.

\begin{figure*}
\begin{center}
\includegraphics[angle=270,width=5.93in]{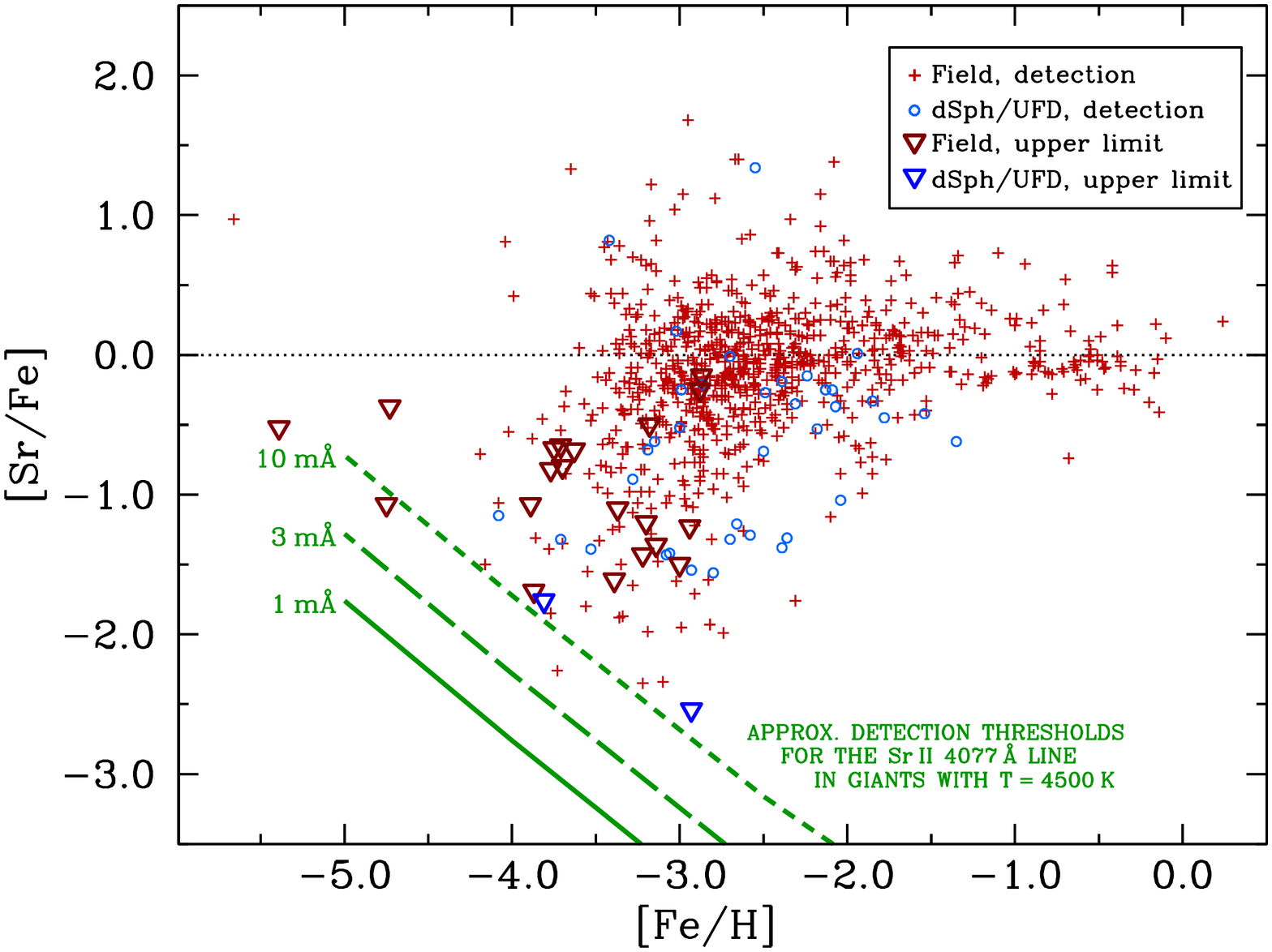}
\end{center}
\caption{
\label{srplot}
[Sr/Fe] ratios in field stars and dwarf galaxies.
Approximate equivalent width detection thresholds 
for the Sr~\textsc{ii} 4077\,\AA\ line 
in a cool giant atmosphere, assuming LTE,
are shown by the set of green curves.
The dotted line marks the solar [Sr/Fe] ratio.
Literature sources for the data are listed in the
caption to Figure~\ref{srbaplot}.
 }
\end{figure*}

\begin{figure*}
\begin{center}
\includegraphics[angle=270,width=5.93in]{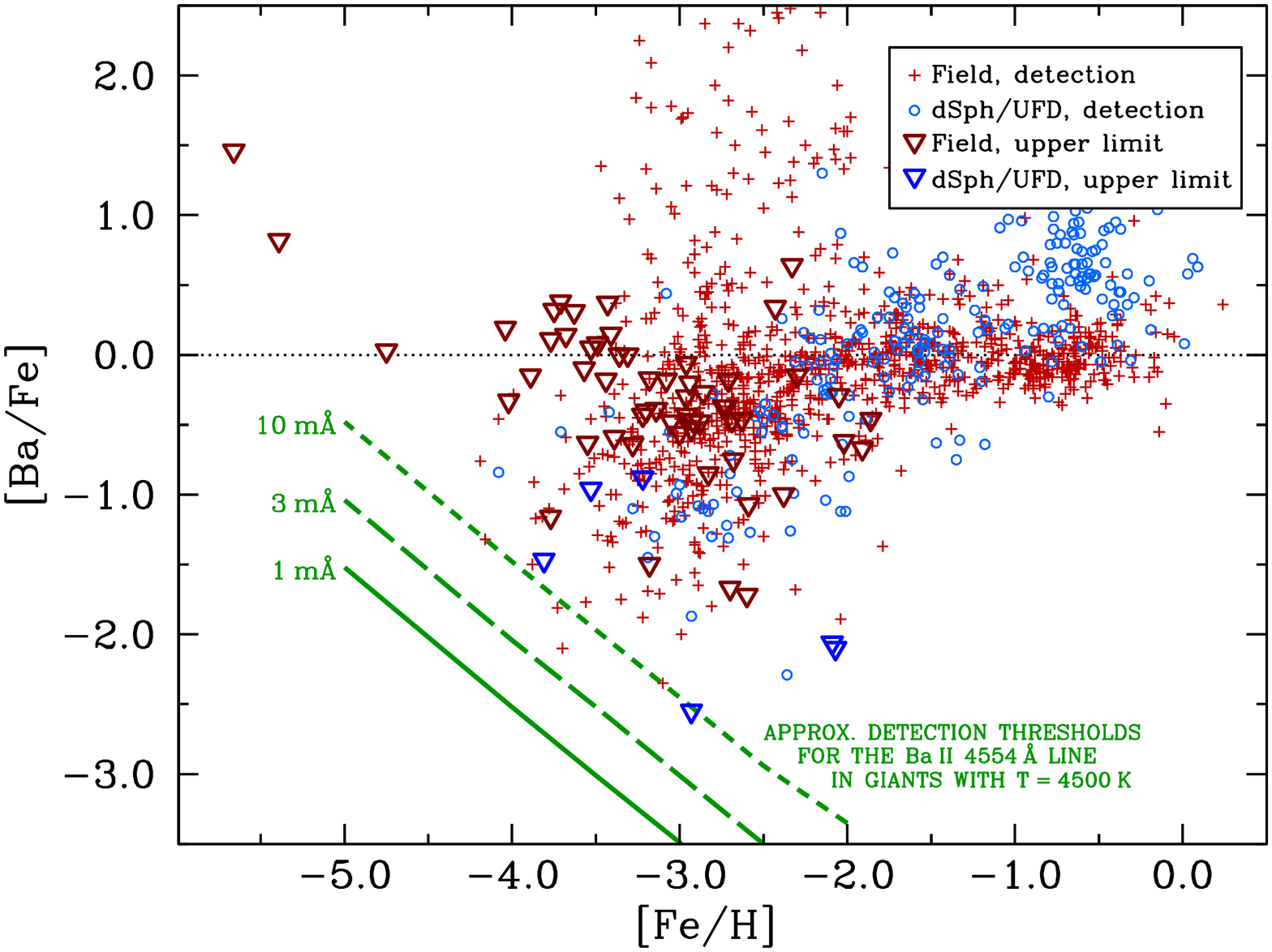}
\end{center}
\caption{
\label{baplot}
[Ba/Fe] ratios in field stars and dwarf galaxies.
Approximate equivalent width detection thresholds 
for the Ba~\textsc{ii} 4554\,\AA\ line 
in a cool giant atmosphere, assuming LTE,
are shown by the set of green curves.
The dotted line marks the solar [Ba/Fe] ratio.
Literature sources for the data are listed in the
caption to Figure~\ref{srbaplot}.
 }
\end{figure*}

Figure~\ref{srbaplot} illustrates the [Sr/H] and [Ba/H] ratios
found in this sample of stars.
Both detections and upper limits are included,
although not all studies report upper limits 
for non-detections.
The upper limits cited in the literature are frequently, but not always,
3$\sigma$ upper limits.
Strontium and barium abundances have not always been reported
together, however, so 
Figures~\ref{srplot} and \ref{baplot} display the
[Sr/Fe] and [Ba/Fe] ratios as a function of [Fe/H].

I have made no attempt to correct these literature values to a 
common \loggf\ or solar abundance scale
or to standardize for different treatments of the
Van der Waals damping constants, $C_{6}$.
The transition probabilities of these lines are each well known
to excellent accuracy.
For example, the NIST Atomic Spectral Database 
\citep{kramida12}
grades their accuracy at 10\% (0.04~dex) or better, 
and the variations in the
\loggf\ values adopted in the 
stellar abundance literature generally mirror this.
Differences in the accepted solar abundances of these
elements are also small, 
varying by $\leq$~0.05~dex among frequently-cited
reviews \citep{anders89,lodders03,asplund09}.
Such differences are negligible for our purposes.

For consistency, all abundances considered in this study are
based on one-dimensional model atmospheres assuming that
local thermodynamical equilibrium (LTE) holds.
Sr~\textsc{ii} and Ba~\textsc{ii} are the dominant 
species in the line-forming layers of FGK-type stars,
so small departures from Saha ionization equilibrium will
have little impact on the derived abundances.
The resonance lines of Sr~\textsc{ii} and Ba~\textsc{ii} 
may be driven out of LTE population equilibrium (non-LTE) by 
underpopulating the lower levels and
overpopulating the upper levels relative to their 
LTE Boltzmann equilibria.
Both Sr~\textsc{ii} and Ba~\textsc{ii} behave similarly 
because of their similar electronic structures.
Such departures are predicted to have a moderate impact
on the derived abundances,
and they are dependent on a variety of factors including 
the temperature, gravity, [Sr/H] or [Ba/H], and the 
collisional cross sections for hydrogen and electrons.
At low metallicity, for a given [Sr/H] or [Ba/H]
the resonance line profile calculated under non-LTE conditions is
weaker than that calculated assuming LTE.
Abundance differences up to $+$0.3~dex or $+$0.4~dex 
are predicted when calculated assuming non-LTE 
relative to the LTE case
for most of the temperature and abundance ranges 
of interest here
\citep{belyakova97,mashonkina99,short06,andrievsky09,andrievsky11}.
While these differences certainly are not negligible,
for low [Sr/H] and [Ba/H]
they will generally raise the abundances and detection thresholds
(Section~\ref{detection}) together. 
This offset does not affect the conclusions of the present study.

\section{Detection Thresholds}
\label{detection}

I have computed approximate thresholds for detecting 
the stronger of the two resonance lines of each species,
Sr~\textsc{ii} 4077\,\AA\ and Ba~\textsc{ii} 4554\,\AA,
in representative model atmospheres ranging from 
cool giants 
(\teff~$=$~4500~K, \logg~$=$~0.5, and \vt~$=$~2.0~\kmsec),
to stars on the lower giant branch
(\teff~$=$~5500~K, \logg~$=$~3.5, and \vt~$=$~1.5~\kmsec),
to warm turn-off stars
(\teff~$=$~6400~K, \logg~$=$~4.0, and \vt~$=$~1.3~\kmsec).
These calculations are made using models interpolated from the
$\alpha$-enhanced grid of ATLAS9 model atmospheres
\citep{castelli03} and the latest version of the
analysis code MOOG \citep{sneden73}
with updates described in \citet{sobeck11}.
The overall metallicity of the model has little effect
($\pm$~0.07~dex) on the calculated limits of [Sr/H] or [Ba/H].

The continuous opacity is lower in cooler giants than
in warm turn-off stars, so absorption lines will be
easier to detect in cooler stars for a constant [Sr/H] or [Ba/H].
In warmer stars, the threshold levels increase.
For example, 
the 3\,m\AA\ detection threshold for
the Sr~\textsc{ii} 4077\,\AA\ line in the 
model with 
\teff~$=$~4500~K is [Sr/H]~$\approx -$6.2.
In the model with 
\teff~$=$~5500~K, the same 3\,m\AA\ detection threshold is
[Sr/H]~$\approx -$4.5, while in the model with
\teff~$=$~6400~K it is [Sr/H]~$\approx -$3.8.
Similarly, the 3\,m\AA\ detection threshold for
the Ba~\textsc{ii} 4554\,\AA\ line in the model with 
\teff~$=$~4500~K is [Ba/H]~$\approx -$6.0.
In the model with
\teff~$=$~5500~K it is [Ba/H]~$\approx -$4.3, while in the model with
\teff~$=$~6400~K it is [Ba/H]~$\approx -$3.5.
The 10\,m\AA, 3\,m\AA, and 1\,m\AA\
detection thresholds for cool giants
are illustrated by three sets of lines
in Figures~\ref{srbaplot}--\ref{baplot}.
These represent the lowest levels of strontium or barium
that could be detected under favorable circumstances.

The Ba~\textsc{ii} 4554\,\AA\ line is broadened by the energy level
shifts from different isotopes of barium, and 
the levels of the two naturally-occurring
odd-$A$ isotopes are further broadened by 
hyperfine splittings.
For weak lines on the linear portion of the curve of growth, 
neglecting this effect has no significant impact on the 
derived abundances.
This is illustrated 
by the low barium abundance cases
in Figure~1 of \citet{mcwilliam98}.

\section{Strontium and Barium in Globular Clusters}
\label{gc}

Heavy element abundance trends in globular cluster stars generally reflect
the patterns in field stars at comparable metallicities.
Globular clusters belonging to the Milky Way and its system
of dwarf galaxies all have mean metallicities 
[Fe/H]~$> -$2.6 or so.
At these metallicities, the highly-enriched and highly-deficient 
extremes of neutron-capture enrichment found in more metal-poor
Galactic halo stars are generally muted
in both field stars and globular clusters
(e.g., \citealt{gratton04}).
Heavy elements are not always studied in late-type stars
in globular clusters.
In studies using modern instrumentation
(since about 1990), no upper limits have been 
published indicating any element, X, heavier than the iron group has
[X/Fe]~$< -$1 in globular clusters
(see, e.g., the compilation by \citealt{pritzl05}).
Some clusters show an internal star-to-star dispersion in 
[X/H] or [X/Fe], but they are always present
(e.g., \citealt{norris95,sneden97,yong08,roederer11}).
The heavy elements are detected 
even in extreme outer halo clusters that reside at
great distances, like Pal~3, Pal~4, Pal~14, or \mbox{NGC~2419}
\citep{koch09,koch10,caliskan12,cohen12}.
Stars in globular clusters associated with the Sagittarius
(e.g., \citealt{brown99})
and Fornax \citep{letarte06} dwarf galaxies
consistently show the presence of heavy elements,
as do those associated with the the Large Magellanic Cloud
\citep{johnson06,mucciarelli08,mucciarelli10}.

There appear to be no globular cluster environments
that were lacking in the elements heavier than the iron group
when the present-day stars were forming.
Since these elements are always found in $\sim$~solar ratios
in globular clusters, they are not shown in the figures.

\section{Strontium and Barium in Field Stars and Dwarf Galaxies}
\label{plots}

Figure~\ref{srbaplot} shows the [Sr/H] and [Ba/H] ratios in 
metal-poor field stars and dwarf galaxies,
and Figures~\ref{srplot} and \ref{baplot} 
show the [Sr/Fe] and [Ba/Fe] ratios as a function of [Fe/H].
These figures reveal that
the number of stars where strontium and barium
are examined but not detected (i.e., upper limits are reported)
constitute only a small fraction of 
all metal-poor stars that have been studied.
Figures~\ref{srplot} and \ref{baplot} demonstrate that
many stars show subsolar [Sr/Fe] and [Ba/Fe] ratios
at metallicities below [Fe/H]~$= -$2.5.
This, of course, has been found repeatedly by many of the studies
whose results are incorporated into the present sample.

The stars of most interest for this study are those
in the lower left corner of Figure~\ref{srbaplot} with
the lowest [Sr/H] and [Ba/H] ratios.
At present, there are no upper limits on [Sr/H] or [Ba/H]
for field stars that are lower than the lowest levels of detection.
\mbox{HE~1116$-$0634}, studied by \citet{hollek11},
shows detectable Sr~\textsc{ii} and Ba~\textsc{ii} lines and
a [Sr/H] ratio ($-$5.99~$\pm$~0.15) 
lower than any other stars studied at present.
Sr~\textsc{ii} and Ba~\textsc{ii} lines are also detected in
\mbox{HE~0302$-$3417}, also studied by \citeauthor{hollek11},
which has 
a [Ba/H] ratio ($-$5.80~$\pm$~0.15) 
lower than any other stars studied at present.
Both of these stars are extremely cool (4400~K),
facilitating the detection of these weak lines.

In the dwarf galaxy sample, 
there are two stars with upper limits on [Sr/H] and [Ba/H] 
that are almost as low as the lowest detections in field stars,
Star~119 in Draco \citep{fulbright04} and
Star~1020549 in Sculptor \citep{frebel10b}.
Four field stars,
\mbox{CS~30336--049} \citep{lai08,yong12},
\mbox{BS~16084--160} \citep{aoki05},
\mbox{CS~22968--014} \citep{mcwilliam95,mcwilliam98,francois07},
and
\mbox{CS~30325--094} \citep{aoki05,francois07}
show detectable Sr~\textsc{ii} and Ba~\textsc{ii} lines and
[Sr/H] and [Ba/H] ratios comparable
to the lower limits found in the two dwarf galaxy stars.
These stars can all be found in the lower left corner of 
Figure~\ref{srbaplot} with 
[Sr/H] and [Ba/H]~$= -$5.5~$\pm$~0.3.

Several independent investigations 
by \citet{shetrone03}, \citet{geisler05}, and \citet{kirby12}
have detected Sr~\textsc{ii} and Ba~\textsc{ii} lines in
numerous other stars in Sculptor
($M_{V} = -$11.1),
including one star with [Fe/H]~$= -$4.0 \citep{tafelmeyer10}.
Draco ($M_{V} = -$8.8) is not completely devoid of
heavy elements, either; 
\citet{shetrone01} and \citet{cohen09} have detected
Sr~\textsc{ii} and Ba~\textsc{ii} lines in other stars in Draco.
Stars~2 and 3 in Hercules ($M_{V} = -$6.6)
show low upper limits,
[Ba/H]~$< -$4.2
([Ba/Fe]~$< -$2.1; \citealt{koch08}).
Sr~\textsc{ii} and Ba~\textsc{ii} lines have been detected 
in other stars in Hercules
(\citealt{francois12,koch12}; A.\ Koch, 2012, private communication),
so Hercules is also not completely lacking heavy elements.

\citet{frebel12} point out that a galaxy like Draco
is luminous enough to have possibly been assembled
from several ``first'' galaxies, 
some of which may not have experienced any 
enrichment of elements heavier than the iron group.
Such galaxies are not found among the surviving UFD galaxies
whose chemistry has been studied.
With the exception of Ba~\textsc{ii} in Segue~1,
Sr~\textsc{ii} and Ba~\textsc{ii} lines have been detected in 
giants in each of the lowest luminosity
dwarf galaxies studied,
Leo~IV ($M_{V} = -$5.8), 
Ursa Major~II ($M_{V} = -$4.2), 
Coma Berenices ($M_{V} = -$4.1), and 
Segue~1 ($M_{V} = -$1.5) 
\citep{frebel10a,norris10b,simon10}. 
All galaxies examined to date show the
presence of heavy elements.

Both Sr~\textsc{ii} resonance lines have been detected
in the most iron-poor star known,
\mbox{HE~1327$-$2326} \citep{frebel05}.
The three other known iron-poor stars with [Fe/H]~$< -$4.5 yield only
upper limits on [Sr/H], but these upper limits
are all significantly lower than the [Sr/H] ratio
found in \mbox{HE~1327$-$2326}.
This cosmic dispersion at the lowest levels of [Fe/H] 
was first pointed out by \citet{aoki06} when less data were available, 
and it still holds true with current data.
The [Sr/H] limits in \mbox{HE~0107$-$5240} \citep{christlieb04}
and \mbox{HE~0557$-$4840} \citep{norris07}
are among the lowest found for any star,
with the exception of \mbox{HE~1116$-$0634} 
as noted above.
The [Sr/H] upper limit given by \citet{caffau12} for 
\mbox{SDSS~J102915$+$172927},
[Sr/H]~$< -$5.1, is a factor of a few higher than these stars.
Nevertheless, the detection of Sr~\textsc{ii} lines 
in \mbox{HE~1327$-$2326}
indicates that at least one mechanism to 
produce elements heavier than the iron group
can operate at the extremely low metallicities
of the stars that enriched the most iron-poor stars.
Ba~\textsc{ii} lines have 
not been detected in any of these stars, 
and the present
upper limits are not low enough to be of great interest.

\section{Nucleosynthesis of Strontium and Barium}
\label{nucleo}

Neutron-capture reactions
are the only known mechanisms for production of elements heavier than
the second neutron-capture peaks (130~$\lesssim A \lesssim$~140).
Charged particle reactions
may be able to produce elements near the first peaks,
like strontium, but they are unable to produce 
elements at or beyond the second peaks, including barium, 
as shown by the calculations of, e.g., \citet{farouqi10}.
The presence of barium and any heavier elements,
including europium, indicates the operation of some kind
of neutron-capture reaction.

Although the abundances of strontium and barium are
considerably lower in halo stars 
than the abundances of lighter elements like
magnesium, calcium, or iron,
the current data indicate that these heavy elements
may be found in nearly every star.
This raises the intriguing prospect that
at least one kind of neutron-capture reaction
operates as frequently as
the nucleosynthesis mechanisms that produce the lighter elements
in the early Universe.
The yields of the heavy elements must certainly be variable
and decoupled from the production of magnesium and iron
to explain the observed
small dispersion in [Mg/Fe]
and large dispersion in [Eu/Fe]. 
The mass ranges of the supernovae that provide this enrichment must also 
play an important role, since there is evidence that some stars may have
been enriched by very few or just one massive 
supernova (e.g., \citealt{simon10}).
Detailed chemical evolution models to test this scenario
are beyond the goals of the present study.

\section{Improving the Observations}
\label{improving}

The detection thresholds indicate that, in principle, there
is still room for improvement in constraining the
upper limits on [Sr/H] and [Ba/H] in cool giants.
Using the relationships between the equivalent width, signal-to-noise
(S/N), and spectrograph parameters given by
\citet{cayrel88} or
\citet{frebel08}, 
it is possible to quantify how much improvement could be expected.
The workhorse high-resolution echelle spectrographs on
the largest optical telescopes are MIKE on the Magellan-Clay Telescope 
\citep{bernstein03},
UVES on the Very Large Telescope \citep{dekker00},
HRS on the Hobby-Eberly Telescope \citep{tull98},
HDS on the Subaru Telescope \citep{noguchi02}, and
HIRES on the Keck~I Telescope \citep{vogt94}.
For standard high-resolution settings 
($R \sim$~30,000 to 50,000) on these instruments,
3$\sigma$ upper limits at S/N~$\sim$~50 pixel$^{-1}$ 
are approximately 3\,m\AA\ to 5\,m\AA\
at the 
Sr~\textsc{ii} and Ba~\textsc{ii}
resonance lines.

Obtaining spectra of this quality of cool giants
in distant dwarf galaxies is challenging since 
Segue~1, Ursa Major~II, 
Coma Berenices, Draco, Hercules and Leo~IV are located at distances of
23~$\pm$~2~kpc \citep{belokurov07},
30~$\pm$~5~kpc \citep{zucker06},
44~$\pm$~4~kpc \citep{belokurov07}, 
76~$\pm$~6~kpc \citep{bonanos04}, 
132~$\pm$~12~kpc \citep{coleman07}, and
154~$\pm$~5~kpc \citep{moretti09}, respectively.
For example, \citet{fulbright04} integrated 10~h with
HIRES on Star 119 in Draco to obtain
S/N~$\sim$~35 pixel$^{-1}$
at the Ba~\textsc{ii} resonance lines
and 4.2~h to obtain S/N~$\sim$~7 pixel$^{-1}$
at the Sr~\textsc{ii} resonance lines.

Additional higher-excitation lines of Ba~\textsc{ii} 
are found at redder wavelengths (5853, 6141, and 6496\,\AA).
These lines are intrinsically weaker than the resonance
Ba~\textsc{ii} lines in cool stars, but they are
useful because of the
increased stellar flux and S/N attainable
at these wavelengths
in comparable exposure times.
Some of the results shown in Figure~\ref{baplot}
(e.g., stars in the Hercules dwarf; \citealt{koch08})
are derived from these lines.

Fortunately,
many of the field giants shown in Figures~\ref{srbaplot}--\ref{baplot}
are within $\approx$~10~kpc of the Sun, so high S/N ratios
in the blue spectral region are attainable.
With deliberate effort to achieve higher S/N ratios 
in the blue (50--100 pixel$^{-1}$),
levels of strontium and barium lower by factors of two to three
could be detected.
Should giants exist with even lower [Sr/H] or [Ba/H] ratios,
the present suite of spectrographs on 6--10~m class telescopes
is capable of identifying them.

\section{Conclusions}
\label{conclusions}

The main result of this study is that no
metal-poor stars have yet been found with
sufficiently low limits on 
[Sr/H] or [Ba/H] to suggest their 
birth environment had not been enriched
by elements heavier than the iron group.
Sr~\textsc{ii} and Ba~\textsc{ii} lines 
are always detected in 
cool globular cluster and field stars
when studied with high-quality observations.
A few stars in some of the low-luminosity
dwarf galaxies may have been born in 
the regions most lacking
in heavy elements,
but strontium and barium have been detected in
at least a few stars in all dwarf galaxies yet studied.

The identification of stars with
unusually low [Sr/H] or [Ba/H] would, of course, 
be of great interest.
Current upper limits can be improved by 
observational campaigns dedicated to
improving the S/N ratios at the blue wavelengths
where the Sr~\textsc{ii} and Ba~\textsc{ii} resonance lines are found.
In cool giants,
upper limits on [Sr/H] and [Ba/H]
that are better by factors of two to three
are attainable with current instruments.

One goal moving forward is to test whether 
all regions where low-mass stars formed 
in and around the halo of the Milky Way 
have experienced at least
minimal amounts of enrichment with elements heavier than the iron group.
Whatever the outcome, this will have profound
implications for characterizing the
frequency and environmental influence of the 
astrophysical sites of heavy element production.

\acknowledgments

The lively discussion among the
participants of the Nuclei in the Cosmos XII satellite workshop
on \rpro\ nucleosynthesis and J.\ Cowan's insightful questions 
served as my inspiration for writing this paper.
I offer my sincerest appreciation to J.\ Cowan, A.\ Koch, 
A.\ McWilliam, G.\ Preston, and D.\ Yong for 
their comments on earlier versions of the figures and manuscript.
I also thank A.\ Koch for sharing results in advance of publication
and the anonymous referee for offering helpful suggestions.
This research has made use of NASA's
Astrophysics Data System Bibliographic Services,
the arXiv pre-print server operated by Cornell University,
the SIMBAD and VizieR databases hosted by the
Strasbourg Astronomical Data Center,
the Stellar Abundances for Galactic Archaeology (SAGA) Database
\citep{suda08}, and
A.\ Frebel's compilation of abundances in field and dwarf galaxy stars
\citep{frebel10c}.
I am grateful for support from the Carnegie Institution for Science
through the Carnegie Fellowship.

\appendix
\section{Additional Comments on the Literature Sample}
\label{appendix}

Repeat observations of the same star by different investigators
are quite common.
In general I have adopted the results derived from the 
higher quality spectra, and
I have avoided mixing abundance ratios reported by 
different investigators for the same star.
There are two cases where one study reported a detection of
Sr~\textsc{ii} but not Ba~\textsc{ii}, but another study
reported a detection of Ba~\textsc{ii} but not Sr~\textsc{ii}.
For these stars, 
\mbox{G64--37} \citep{aoki09,ishigaki10} and
\mbox{HD~184499} \citep{fulbright00,mishenina01},
I have included both detections.

For one star, \mbox{CS~22949--048}, I include the 
non-LTE abundances of [Sr/H] and [Ba/H].
\citet{mcwilliam98} and \citet{lai07} each cite upper limits
on [Ba/Fe] in this star, 
but \citet{andrievsky11} report a detection of Ba~\textsc{ii} and an 
abundance derived from (only) non-LTE calculations.
The [Sr/Fe] and [Ba/Fe] ratios in this star are subsolar but otherwise
unremarkable.

\citet{lai07} reported abundances for \mbox{CS~22962--006}.
This star was misidentified and should instead be
classified as a white dwarf (D.\ Lai, 2012, private communication).
Consequently this star is not included in the present sample.

\end{document}